\documentclass[11pt,a4paper]{article}
\usepackage[scale=0.8]{geometry}

\title{Are trading invariants really invariant? Trading costs matter}

\usepackage{authblk}

\author[1,7]{Fr\'ed\'eric Bucci \footnote{Corresponding author: frederic.bucci@sns.it}}
\author[2,3]{Fabrizio Lillo}
\author[4,5,7]{Jean-Philippe Bouchaud}
\author[4,6,7]{Michael Benzaquen}
{\affil[1]{\small  Scuola Normale Superiore di Pisa, Piazza dei Cavalieri 7, 56126 Pisa, Italy}
\affil[2]{Department of Mathematics, Universit\`a di Bologna, Piazza di Porta San Donato 5, 40126 Bologna, Italy}
\affil[3]{CADS,  Human  Technopole,  Milan,  Italy}
\affil[4]{Capital Fund Management, 23 rue de l'Universit\'e, 75007, Paris, France}
\affil[5]{CFM-Imperial Institute of Quantitative Finance, Department of Mathematics, Imperial College, 180 Queen's Gate, London SW7 2RH}
\affil[6]{Ladhyx UMR CNRS 7646 \& Department of Economics, Ecole polytechnique, 91128 Palaiseau Cedex, France}
\affil[7]{Chair of Econophysics and Complex Systems, Ecole polytechnique, 91128 Palaiseau Cedex, France}}

\date{\today}

\usepackage{amssymb}
\usepackage{graphicx}
\usepackage{amsmath}
\usepackage{amsfonts}
\usepackage{geometry}
\usepackage{bm}
\usepackage{breqn}
\usepackage{textcomp}
\usepackage{xcolor}
\usepackage{subfig}
\usepackage{amsmath}
\usepackage{amssymb}
\usepackage{geometry}
\usepackage{mathtools}
\usepackage{tabularx}

\newcommand{\avg}{\mathbb{E}}

\newcommand{\ri}{{\mathcal{R}}}

\begin{document}
\maketitle

\vspace{-1cm}

\abstract{We revisit the trading invariance hypothesis recently proposed by Kyle and Obizhaeva \cite{Kyle1} by empirically investigating a large dataset of bets, or metaorders, provided by ANcerno. The hypothesis predicts that the quantity $I:=\ri/N^{3/2}$, where $\ri$ is the exchanged risk (volatility $\times$ volume $\times$ price) and $N$ is the number of bets, is invariant. We find that the $3/2$ scaling between $\ri$ and $N$ works well and is robust against changes of year, market capitalisation and  economic sector. However our analysis clearly shows that $I$ is not invariant. We find a very high correlation $R^2>0.8$ between $I$ and the total trading cost (spread and market impact) of the bet. We propose new invariants defined as a ratio of $I$ and costs and find a large decrease in variance. We show that the small dispersion of the new invariants is mainly driven by (i) the scaling of the spread with the volatility per transaction, (ii) the near invariance of the distribution of metaorder size and of the volume and number fractions of bets across stocks.}

\tableofcontents

\section{Introduction}

Finding universal scaling laws between trading variables is highly valuable to make progress in our understanding of financial markets and market microstructure. In the wake of these discoveries, Kyle and Obizhaeva posit a \textit{trading invariance principle} that must be valid for  a \textit{\textit{bet}}, theoretically defined as a sequence of orders with a fixed direction (buy or sell) belonging to a single trading idea \cite{Kyle1,Kyle2}. This principle supports the existence of a universal invariant quantity $I$ -- expressed in dollars, independent of the asset and constant over time -- which represents the average cost of a single \textit{\textit{bet}}. In particular, taking the share price  $P$ (in dollars per share), the square daily volatility $\sigma_{\mathrm{d}}^2$ (in $\%^2$ per day), the total daily amount traded with \textit{\textit{bet}s} $V$ (shares per day) and the average volume of  an individual \textit{\textit{bet}} $Q$ (in shares) as relevant variables,  dimensional analysis suggests a relation of the form:
\begin{equation}
\frac{PQ}{I}=f\left(\sigma_{\mathrm{d}}^2\frac{Q}{V}\right) \ ,
\end{equation}
where $f$ is a dimensionless function. Invoking the Modigliani-Miller capital structure irrelevance principle yields $f(x)\sim x^{-1/2}$, which implies up to a numerical factor that: 
\begin{equation}
\label{eq:1}
I=\frac{\sigma_{\mathrm{d}} P Q^{3/2}}{V^{1/2}}:=\frac{\ri}{N^{3/2}} \ ,
\end{equation}
where $\ri:= \sigma_{\mathrm{d}}PV$  measures  the total dollar amount of risk traded per day (also referred to as total exchanged risk or trading activity) while $N:=V/Q$ represents the  number of daily \textit{\textit{bet}s}  for a given  contract. Notwithstanding, the 3/2 law can be interpreted with different degrees of universality as discussed in \cite{Benzaquen}: no universality (the 3/2 holds for some contracts only), weak universality (the 3/2 holds but with a non-universal value of $I$) and strong universality (the 3/2 holds and $I$ is constant across assets and time).\footnote{ Note that here we only explore the daily level, time does not mean the same thing as in \cite{Benzaquen} where we varied the time intervals over which the variables were computed.}

Let us stress that identifying an elementary \textit{bet} in the market is not a straightforward task. Theoretically, a bet is defined as a trading idea typically executed in the market as many trades over several days. As suggested by Kyle and Obizhaeva in their original work \cite{Kyle1},  metaorders, i.e. a bundle of orders corresponding to a single trading decision typically traded incrementally through a sequence of child orders, can be considered a proxy of these \textit{bet}s; in this work we will make use of such an  approximation and use the words `bet' or `metaorder' indifferently.  Beyond the subtleties in the {bet}'s definition, there has been in the past few years empirical evidence that the scaling law discussed above matches patterns in financial data, at least approximately. The  3/2-law was empirically confirmed by Kyle and Obizhaeva using portfolio transition data  related to rebalancing decisions made by institutional investors and executed by brokers \cite{Kyle1}.  Andersen \textit{et al.} \cite{Andersen} reformulated suitably the \textit{trading invariance hypothesis} at the single-trade level and  showed that the equivalent version of Eq.~\eqref{eq:1} in such a setting holds remarkably well using public trade-by-trade data relative to the E-mini S\&P 500 futures contracts.  Benzaquen \textit{et al.}  \cite{Benzaquen} substantially extended these empirical results showing that the $3/2$-law holds very precisely across 12 futures contracts and 300 single US stocks, and across a wide range of time scales. Recently, Pohl \textit{et al.} \cite{Pohl}  provided additional empirical evidence that the intriguing $3/2$-law holds on trades data from the NASDAQ stock exchange.

Notwithstanding, empirical data at the single transaction scale -- see in particular \cite{Benzaquen} -- revealed that while the 3/2 law is very robust, the invariant $I$ is actually quite far from invariant, as it varies from one asset to the other and across time, thus in favour of the \emph{weak universality} degree. Note that this is consistent with the idea that a universal invariant with dollar units would be quite strange, given that the value of the dollar is itself time-dependent. Benzaquen \textit{et al.} \cite{Benzaquen} showed that a more suitable candidate for an invariant was actually the dimensionless $\mathcal I:=I/\mathcal C$ where $\mathcal C$ denotes the spread trading costs. 

Yet, single transactions are typically not the same as single bets. Large and medium sized orders are typically split in multiple transactions and traded incrementally over long periods of time. Empirically market data do not allow to infer the trading decision and to link different transactions to a single execution.\footnote{In fact, for example, Kyle and Obizhaeva tackled this problem investigating a proprietary dataset of portfolio transitions.} In order to test the trading invariance hypothesis at the bet level and its relation with trading costs, it is necessary to have a dataset of market-wide (i.e. not from a single institution) metaorders. 
 
This is precisely the aim of the present paper, which leverages on a heterogeneous dataset of metaorders extracted from the ANcerno database.\footnote{ANcerno Ltd. (formerly the Abel Noser Corporation) is a widely recognised consulting firm that works with institutional investors to monitor their equity trading costs. Its clients include many pension funds and asset managers. In \cite{Kyle1} the authors claim that the ANcerno database includes more orders than the data set of portfolio transitions they used in their work.  From a preliminary research Albert S. Kyle and Kingsley Fong  found that proxies for \textit{bet}s in ANcerno data have size patterns consistent with the proposed invariance hypothesis discussed in \cite{Kyle1,Kyle2}.} To our knowledge such a thorough analysis at the \textit{bet} level for a wide range of assets is still lacking.  

Our main finding is that, while the $3/2$-law works surprisingly well, the quantity $I$ is not invariant, as pointed out in \cite{Benzaquen}. We show that this quantity is strongly correlated with transaction costs, including spread and impact. Therefore we  introduce new invariants, obtained by dividing $I$ by the cost and we show that these quantities fluctuate very little across stocks and time periods. Finally we show that the observed small dispersion of the new invariants can be connected with three microstructural properties: (i) the linear relation between spread and volatility per transaction; (ii) the near invariance of the metaorder size distribution, and (iii) of the total volume and number fractions of the bets across different stocks.
 
The paper is organized as follows. In section \ref{data} we describe the dataset  collecting trading decisions of institutional investors operating in the US equity market. In section \ref{3su2law} we show that the $3/2$-law holds surprisingly well at the daily level independently of the time period, of the market capitalisation and of the economic sector. In section \ref{tradinv} we compute the invariant $I$ and we argue in favour of \textit{weak universality}.  We propose a more natural definition for a trading invariant that accounts both for the spread and the market impact costs; and we exhibit the microstructural origin of its small dispersion.  Some conclusions and open questions  are presented in section \ref{conclusion}.

\section{Data}
\label{data}
Our analysis relies on a database made available by  ANcerno, a leading
transaction-cost analysis provider ({\tt www.ancerno.com}). Our  dataset counts heterogeneous institutional investors placing large buy or sell orders executed by a broker as a succession of smaller orders belonging  to the same trading decision of a single investor. Our sample includes the period January  2007~--~June 2010 for a total of 880 trading days. Only metaorders completed within at most a single trading day are held. Further, we select stocks belonging to the Russell 3000 index, thereby retaining $\sim$ 8 million metaorders distributed quite uniformly in time and representing $\sim 5\%$ of the total reported market volume, regardless of  market capitalisation (large, mid and small) and economical sectors (basic materials, communications, consumer cyclical and non-cyclical, energy, financial, industrial, technology and utilities).  More details and statistics on the investigated sample are presented in Appendix \ref{app_data}.

\section{The 3/2-law}
\label{3su2law}
Here we  investigate the trading invariance hypothesis at the daily level. The daily timescale choice avoids an elaborate analysis of when precisely each metaorder starts and ends, thereby averaging out all the non-trivial problems related to the  daily simultaneous  metaorders executed on the same asset \cite{Zarinelli}. 
\subsection{Exchanged risk}
From the metaorders executed on the same stock during the same day we  compute the total  exchanged  volume in dollars:
$\sum_{i=1}^N \textrm{p}_i \textrm{v}_i$, where $N$ is the number of daily metaorders per asset in the ANcerno database, $\textrm{v}_i$ and $\textrm{p}_i$ are respectively  the number of shares and the volume weighted average price (vwap) of the $i$-th available metaorder. We then define  the total daily exchanged ANcerno risk per asset as:
\begin{equation}
\ri:=\sum_{i=1}^N \ri_i \ , \quad \textrm{with}\quad \ri_i=\sigma_{\mathrm{d}} \textrm{v}_i \textrm{p}_i \ ,
\end{equation}
\begin{figure}[t!]
\centering
\includegraphics[width=0.88\linewidth]{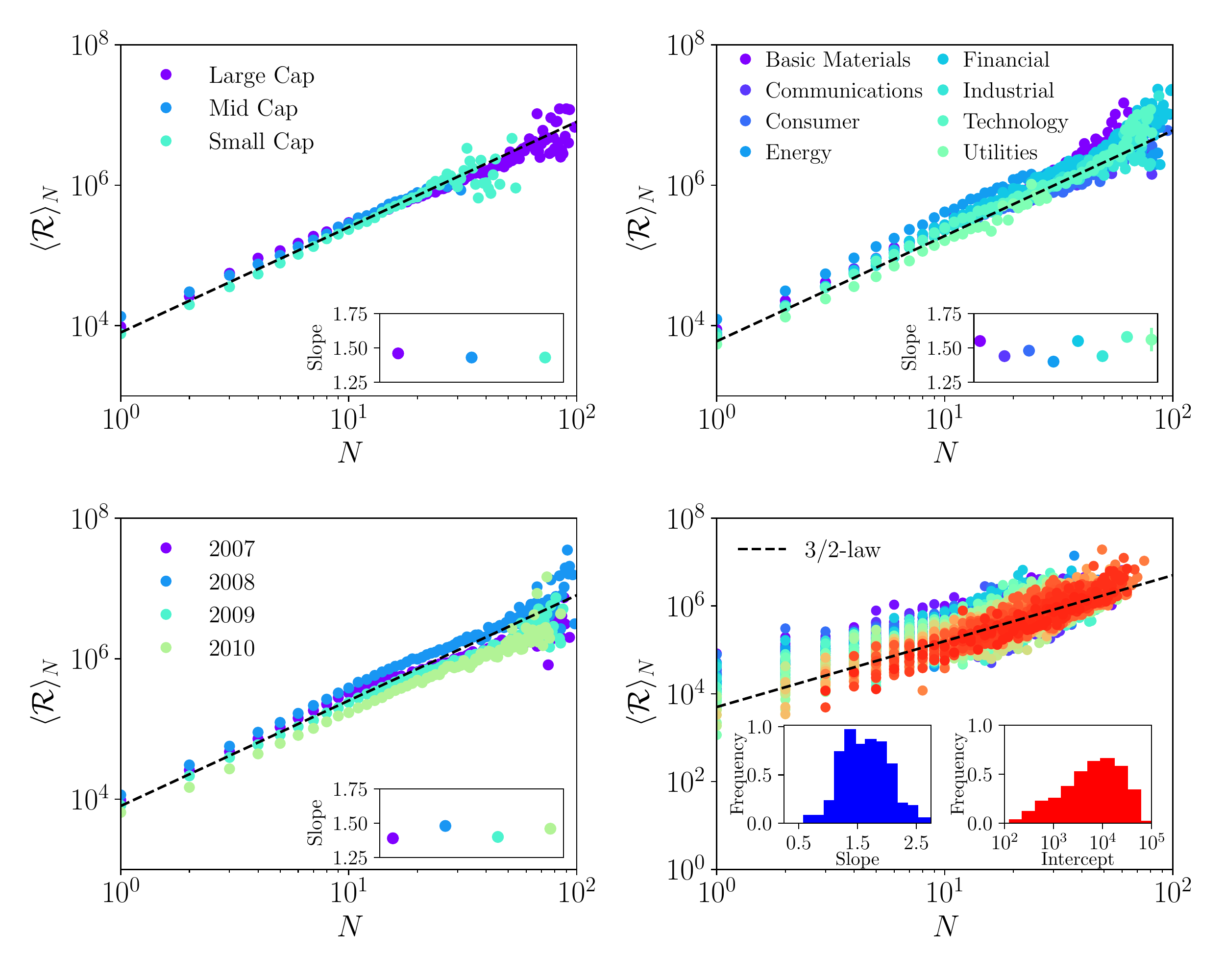}
\caption{Plots of the mean daily exchanged risk  $\langle \mathcal{R} \rangle_N$ as function of the daily number $N$ of metaorders per asset conditional to the  market capitalisation (top left panel),  the  economic sector (top right panel), and  the time period (bottom left panel). The insets show the slopes obtained from linear regression  of the data, firstly averaged respect to $N$ and secondly log-transformed. The bottom right panel shows a plot  of  $\langle \mathcal{R} \rangle_N$ as function of $N$ for a subset of 200  stocks chosen randomly from the pool of around three thousand US stocks: the two insets represent respectively the distribution of the slopes and  of the $y$-intercept, i.e. $\langle I \rangle:=\langle \ri \rangle_N /N^{3/2}$, obtained from linear regression  of the data, firstly averaged respect to $N$ and secondly log-transformed of the data considering each stock separately.}
\label{figure1}
\end{figure}
and  where $\sigma_{\mathrm{d}}$ denotes  the daily volatility per asset,  computed as $\sigma_{\mathrm{d}}=(\textrm{p}_{\textrm{high}}-\textrm{p}_{\textrm{low}})/\textrm{p}_{\textrm{open}}$ from    the high, low,  and open daily prices only.\footnote{We checked that the  results discussed in the present work are still valid using other definitions of the daily volatility and of the price in analogy to what done for example in \cite{Kyle1}. Specifically, the results are still valid when computing $\sigma_{\mathrm{d}}$ with the Rogers-Satchell volatility estimator \cite{Benzaquen,RS} or  as  the monthly averaged daily  volatility, i.e. $\bar \sigma_{\mathrm{d}}=\sum_{m=1}^{25} \sigma_{d,m}$  and/or defining the price $\textrm{p}_i$ as  the closing price of the day before  the metaorder's execution.} The statistical properties of the bets, in terms of their associated risk  $\mathcal{R}_i$ and of their total daily number $N$ per asset are discussed in Appendix \ref{app_data}. The variability of the observables  over several orders of magnitude should  allow to test the 3/2-law quite convincingly.

\subsection{Empirical evidence}
We introduce the mean daily exchanged risk $\langle \mathcal{R} \rangle_N$, where  in general $\langle \bullet \rangle_N:=\avg[\bullet|N]$ denotes in a compact way the average over various days and stocks with a fixed daily number of  metaorders  $N$. As shown in the first three panels of  Fig.~\ref{figure1}  the scaling $\langle \ri \rangle_N \sim N^{3/2}$ holds well independently of the conditioning to  market capitalisation,   economic sector, and  time period. Slight deviations may have different origins but can mostly be attributed to the heterogeneous sample's composition in terms of stocks for each bucket in $N$. The $3/2$-law is also valid  for  individual stocks,  as shown in the bottom right panel of Fig.~\ref{figure1}. The bottom left inset shows that the slopes are clustered around the $3/2$-value. However, the $y$-intercepts of the fitted lines for individual stocks vary substantially  (see the bottom right inset in the bottom right panel of Fig.~\ref{figure1}), which indicates that $I$ is not constant across different stocks.  More empirical insights on the origin of the $3/2$ law are presented in Appendix \ref{app_micro}.


\section{The trading invariant}
\label{tradinv}

The conjecture that the quantity  $\langle I \rangle:=\langle \ri \rangle_N /N^{3/2}$ 
is invariant across different contracts is clearly rejected by the empirical analysis performed in the previous section. Indeed, the quantity $\langle I \rangle$ varies by at least one order of magnitude across different stocks. This result goes  against  the \textit{strong universality} version of the trading invariance hypothesis which states that both the average value $\langle I \rangle$ and the full probability distribution of $I=\ri/N^{3/2}$ should be invariant across products.  Dimensionally $I$ is a cost (i.e. it is measured in dollars) and indeed the trading invariance hypothesis posits that the cost of a bet is invariant. Using the identification of metaorders and bets we can use the ANcerno dataset to estimate the trading cost, including a spread and a market impact component. We will show that $I$ and trading cost are very correlated, and therefore propose new invariants based on their ratio.

\subsection{Trading costs and trading invariants}

 \begin{figure}[b!]
\begin{minipage}{1.0\textwidth}
\centering
\includegraphics[width=0.8\linewidth]{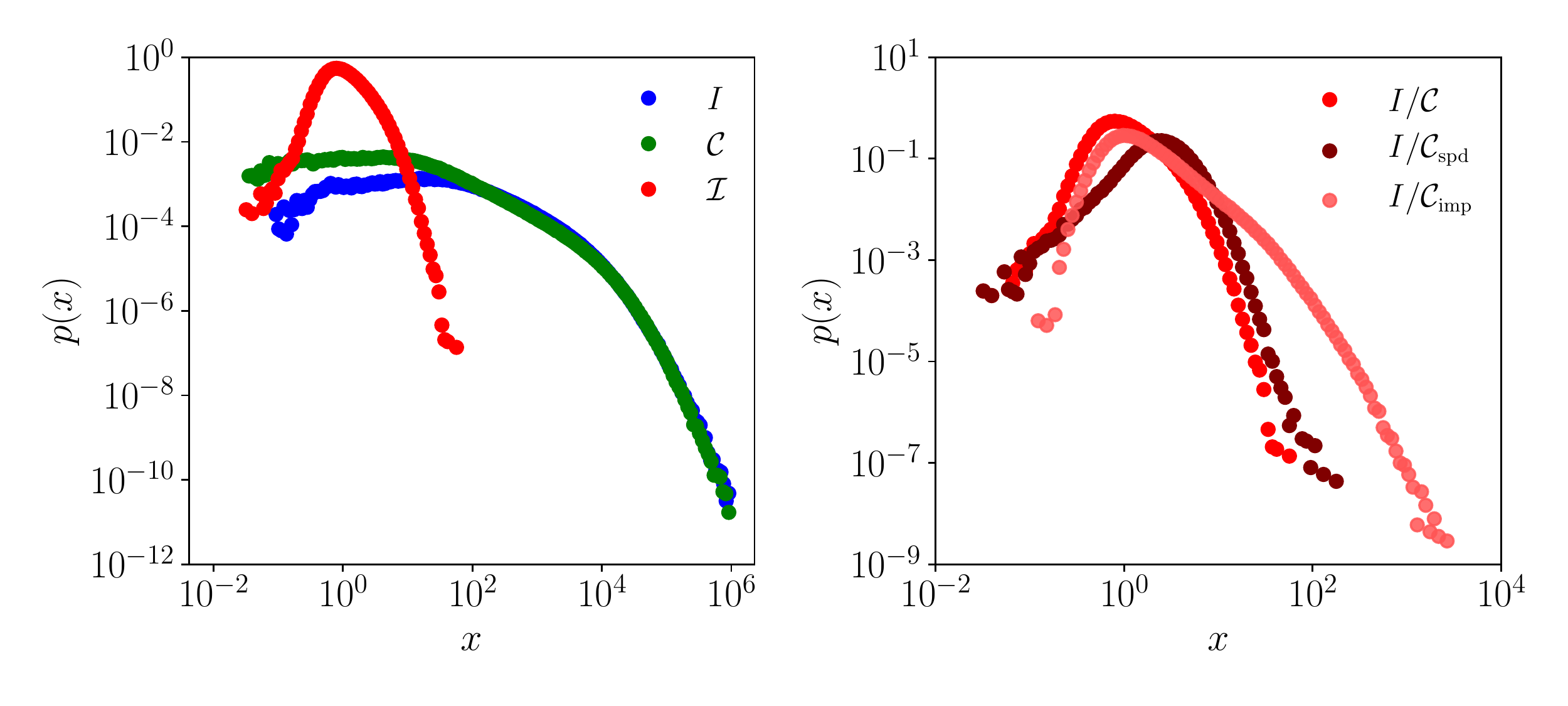}
\end{minipage}
\caption{(Left) Empirical distributions of the KO invariant $I=\mathcal{R}N^{-3/2}$, of the daily average  \textit{\textit{bet}}'s total trading cost $\mathcal{C}$ (using $Y_{\textrm{spd}}=3.5$ and $Y_{\textrm{imp}}=1.5$  in  Eq. \ref{gencost}), and   of the dimensionless invariant $\mathcal{I}:=I/\mathcal{C}$. (Right) Empirical distributions in log-log scale of the KO invariant  $I$ rescaled respectively by the total daily average cost $\mathcal{C}$, by the spread cost $\mathcal{C}_{\textrm{\mathrm{spd}}}$ and by the market impact cost $\mathcal{C}_{\textrm{imp}}$.}
\label{figureICItilde2}
\end{figure}

Trading costs are typically divided into fees/commissions, spread, and market impact. For large orders, like those investigated here, fees/commissions typically account for a very small fraction and therefore we will neglect them. 
We shall however take into consideration both the spread cost  (as was done at the single-trade level in \cite{Benzaquen}) and the market impact cost  computed from the square root law (see \emph{e.g.} \cite{Zarinelli,BToth,Torre,Almgren,Engle,Brokmann, Bucci, Bucci2}). We thus define the average daily {\textit{bet}}'s trading cost as: 
\begin{equation}
\mathcal{C}=\mathcal{C}_{\mathrm{spd}}+\mathcal{C}_{\mathrm{imp}}=Y_{\textrm{spd}}\times\frac{1}{N}\sum_{i=1}^N S\textrm{v}_i +Y_{\textrm{imp}} \times\frac{1}{N} \sum_{i=1}^N \sigma_{\mathrm{d}} \textrm{v}_i \textrm{p}_i \sqrt{\frac{\textrm{v}_i}{V_{\mathrm d}}} := Y_{\textrm{spd}} \times \mathcal{C}^0_{\mathrm{spd}} + Y_{\textrm{imp}}  \times \mathcal{C}^0_{\mathrm{imp}} \ ,
\label{gencost}
\end{equation} 
 with $S$  the average daily spread,\footnote{The daily spread is recovered from a dataset provided by CFM since it is not available in the ANcerno dataset.} $V_{\mathrm d}$ the total daily market volume, $Y_{\textrm{spd}}$ and  $Y_{\textrm{imp}}$  two constants to be determined. The factor $Y_{\textrm{spd}}$ depends, among other things, on the fraction of trades of the metaorder executed with market orders, whereas $Y_{\textrm{imp}}$ only weakly depends on the execution algorithm and is typically estimated to be very close to unity \cite{Zarinelli, BToth,bouchaud2018trades}. The empirical properties of ${\mathcal{C}}^0_{\textrm{\tiny{\mathrm{spd}}}}$ and ${\mathcal{\huge{C}}}^0_{\textrm{\tiny{imp}}}$ and the relative importance of the two terms as a function of the metaorder size are presented in Appendix \ref{app_costs}.
To determine $Y_{\textrm{spd}}$ and  $Y_{\textrm{imp}}$  we perform an ordinary least square regression of the KO invariant $I$ with respect to the daily average cost $\mathcal{C}$ defined for each asset by Eq. \ref{gencost}. We obtain $Y_{\textrm{spd}}\simeq 3.5$, $Y_{\textrm{imp}}\simeq 1.5$ and a coefficient of determination $R^2\simeq 0.8$. These results show that  the original KO invariant is indeed strongly correlated with the trading cost. Since these costs have no a priori reason to be universal, this explains why $I$ is not invariant.

Guided by such results and by the fact that a market microstructure invariant, if any, should be dimensionless, we define new invariants by dividing the original KO invariant $I$ by the cost of trading. Therefore, we consider three different specifications, namely:
\begin{equation}\label{newInv}
\mathcal{I}=\frac{I}{\mathcal{C}}\ ,~~~~~~~~~~~~~~~~~\mathcal{I}_{\mathrm{spd}}=\frac{I}{\mathcal{C}_{\textrm{\tiny{\mathrm{spd}}}}}\ ,~~~~~~~~~~~~~~~\mathcal{I}_{\mathrm{imp}}=\frac{I} {\mathcal{C}_{\textrm{\tiny{imp}}}}\ .
\end{equation}

The left panel of Figure \ref{figureICItilde2} shows the distribution of the original KO invariant $I$ together with that of $\mathcal{I}$, and of the cost $\mathcal{C}$. It is visually quite clear that rescaling by the cost dramatically reduces the dispersion, and that the distribution of $I$ is very similar to that of $\mathcal{C}$, with some deviation for small value. The right panel compares the distribution of $\mathcal{I}$ with that of the other two new invariants. A quantitative comparison is provided in Table \ref{tab:cv}, which reports  the mean, the standard deviation, and the coefficient of variation\footnote{The coefficient of variation is the ratio of standard deviation and mean, an indicator of  distribution `peakedness'.}  (CV) of $I$ and of the three new invariants.  It is clear that, due to the correlation between $I$ and $\mathcal{C}$, the new invariants $\mathcal{I}_{\circ}$ (with $\circ=\textrm{spd}, \textrm{imp}$) have a much smaller CV than $I$. Since the distributions have clear fat tails, we also implemented a robust version of CV obtained by replacing the standard deviation with the mean absolute deviation (MAD), here denoted $\mbox{CV}_{\mathrm{MAD}}$. The table indicates that also in this case the new invariants are much more peaked than $I$. Notice also that with $\mbox{CV}_{\mathrm{MAD}}$ the three new invariants become similar, while with CV the invariant $\mathcal{I}_{\mathrm{imp}}$ is more dispersed.

 \begin{table}[h!]
\caption{Statistics of the different invariants, namely the original KO invariant $I$ (left), and the three new ones rescaled by cost (right). MAD is the mean absolute deviation and CV stands for coefficient of variation.\vspace{-0.5cm}}\label{tab:cv}
\begin{center}
\begin{tabular}{c|c|ccc}
\hline
 &$I\cdot 10^{3}$ ($\$$)&$\mathcal{I}$  & $\mathcal{I}_{\textrm{\mathrm{spd}}}$ & $\mathcal{I}_{\textrm{imp}}$\\
 
\hline
\hline
 mean&6.33&2.20&4.70&7.8\\
st. dev.&11& 1.84 &3.11&12.2  \\
MAD&6.9& 1.25& 2.21&7.56\\
\hline
\hline
CV&1.74&0.84&0.66&1.56
\\
$\mbox{CV}_{\mathrm{MAD}}$&1.09&0.57&0.47&0.97\\
\hline
\end{tabular}
\end{center}
\end{table}

\subsection{Origin of the small dispersion of the new invariants}

\begin{figure}[t!]
\centering
\includegraphics[width=1.0\linewidth]{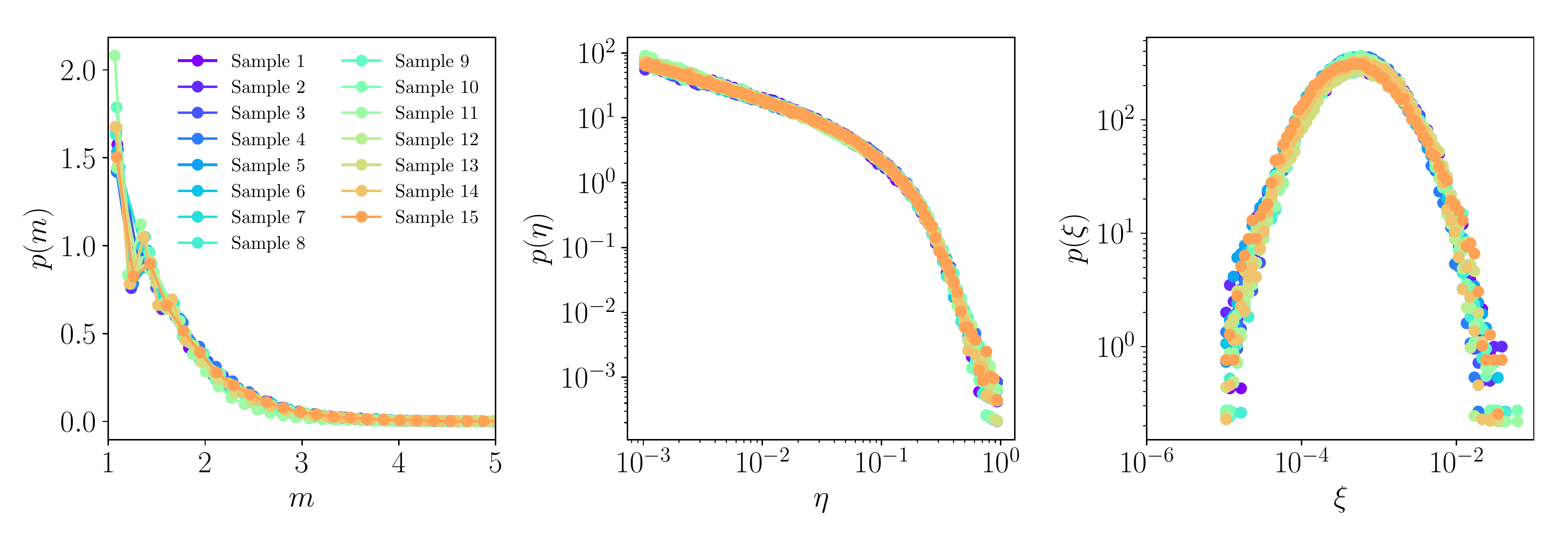}
\caption{Empirical distributions of the  ratio $m=[\textrm{v}^{3/2}]/[ \textrm{v}]^{3/2}$ (left panel), $\eta=V/V_{\mathrm{d}}$ (central panel) and $\xi = N/N_{\mathrm{d}}$ (right pannel), all three computed at the daily level  for each asset: we randomly  group the stocks in equally sized samples and for each of them we compute the empirical distribution respectively of $m$, $\eta$ and $\xi$ finding that they are, to a first approximation, stock independent.}
\label{fig20}
\end{figure}

Here we investigate the origin of the small dispersion of the new invariants.
Let us first consider only the impact cost normalisation only and rewrite $\mathcal{I}_{\mathrm{imp}}$ as:
\begin{equation}
\mathcal{I}_{\mathrm{imp}} =\frac{N\sum_{i=1}^N \sigma_{\mathrm{d}} \textrm{p}_i \textrm{v}_i}{Y_{\textrm{imp}} N^{3/2}(\sigma_{\mathrm{d}} \sum_{i=1}^N \textrm{p}_i \textrm{v}_i \sqrt{\textrm{v}_i/V_{\mathrm{d}}})} \ .
\end{equation}
Using $\textrm{p}_i\simeq \textrm{p}$, for all the metaorders executed in a day on a stock, the above expression simplifies to:
\begin{equation}
\mathcal{I}_{\mathrm{imp}} =\frac{1}{Y_{\textrm{imp}} \sqrt{\eta}} \frac{[\textrm{v}]^{3/2}}{[ \textrm{v}^{\,3/2} ]}=\frac{1}{Y_{\textrm{imp}} m\sqrt{\eta}}  \ ,
\end{equation}
where  $\eta:=V/V_{\mathrm d}$ with $V:=\sum_{i=1}^N \textrm{v}_i$ is the total ANcerno bet volume, $[ \bullet ]$ is a daily average operation per stock, and $m>1$ is the normalized 3/2th moment of $\textrm{v}$ (number of shares of a bet), which depends on the {\it shape} of the distribution of metaorder size. We have checked that $m$ as well as $\eta$  are, to a first approximation, independent of the stock (see left and central panels in Fig.~\ref{fig20})  indicating that the distribution of metaorder size is, to a large degree, universal and that the ANcerno database is representative of the trading across all stocks. These observations  explains why $\mathcal{I}_{\mathrm{imp}}$ is also, to a large degree, stock independent. 

For the total cost normalisation, our understanding of the invariance property relies on the following empirical fact.
The average spread is proportional to the volatility per trade, that is $ S= c\textrm{p}\sigma_{\mathrm{d}} /\sqrt{N_{\mathrm{d}}}$, where $c$ is a stock independent numerical constant, see \cite{bouchaud2018trades,Wyart}.
Indeed, the above arguments taken together show that the dimensionless quantity $\mathcal{I}$ can be written as:
\begin{equation}
\mathcal{I} =  \frac{1}{ Y_{\textrm{spd}} c\sqrt{\xi} + Y_{\textrm{imp}} m\sqrt{\eta} },
\end{equation}
where $\xi:=N/N_{\mathrm{d}}$ is found to be stock independent (see right panel in Fig.~\ref{fig20}). Therefore $\mathcal{I}$ is also stock independent. However, the fact that the CV of $\mathcal{I}$ is less than both those of $\mathcal{I}_{\mathrm{spd}}$ and $\mathcal{I}_{\mathrm{imp}}$ suggests that the Kyle-Obizhaeva ``invariant'' reflects the fact that metaorders are commensurate to the {\it total} cost of trading, including both the spread cost and the impact cost.


\section{Conclusions}
\label{conclusion}

In this work we empirically investigated the \textit{market microstructure invariance hypothesis} recently proposed by  Kyle and Obizhaeva \cite{Kyle1, Kyle2}. Their conjecture is that the expected dollar cost of executing a \textit{bet} is constant across assets and time.  The ANcerno dataset provides a unique laboratory to test this intriguing hypothesis through its available  metaorders which can be treated as a proxy for  \textit{bet}s, i.e. a decision to buy or sell  a quantity of institutional size generated by a specific trading idea.
Let us summarise what we have achieved in this paper:
\begin{itemize}
\item Using \textit{bets} issued for around three thousand stocks, we showed that, at the daily timescale interval, the $N^{3/2}$ scaling law between exchanged risk $\mathcal{R}$ and number of bets is  observed independently of the year,  the economic sector and  the market capitalisation.

\item The trading invariant $I:=\mathcal{R}/N^{3/2}$ proposed by Kyle and Obizhaeva is non-universal: both its average value $\langle I \rangle$ and its whole distribution clearly depend on the considered stocks, in favour of a \textit{weak universality} interpretation. Furthermore, this quantity has dollar units which makes its hypothesised invariance rather implausible. 

\item On the basis of dimensional and empirical arguments, we propose a dimensionless invariant defined as a ratio of $I$ and of the bet's total cost, which includes both spread and market impact costs. We find a variance reduction of more than 50\%, qualitatively traceable to the proportionality between spread and volatility per trade, and the near invariance of the distributions of bet size, of the  volume fraction and number fraction of bets across stocks.
\end{itemize}

Our empirical analysis has allowed to show that the trading invariance hypothesis holds at the \textit{bet} level in a strong sense provided one considers the exchanged risk {\it and} the total trading cost of the \textit{bets}. This is in the spirit of Kyle and Obizhaeva's arguments, but takes into account the fact that transaction costs are both asset and epoch dependent. As anticipated in \cite{Benzaquen}, our results strongly suggest that trading ``invariance'' is a consequence of the endogeneisation of costs in the trading decision of market participants, and has little to do with the Modigliani-Miller theorem. It would actually be quite interesting to investigate other markets such as bond markets, currency markets or futures markets, for which the Modigliani-Miller theorem is totally irrelevant, while trading invariance still holds -- at least at the level of single trades \cite{Benzaquen, Andersen}. Finally, differences in market structure across countries, such as execution mechanisms, fees and regulations could also challenge the validity of the results presented here. 

\section*{Acknowledgments}
We thank Alexios Beveratos, Laurent Erreca, Antoine Fosset, Charles-Albert Lehalle  and Amine Raboun for fruitful discussions. This research was conducted within the \emph{Econophysics \& Complex Systems} Research Chair, under the aegis of the Fondation du Risque, the Fondation de l'Ecole polytechnique, the Ecole polytechnique and Capital Fund Management.

\section*{Data availability statement}
The data were purchased from the company ANcerno Ltd (formerly the Abel Noser Corporation) which is a widely recognised consulting firm that works with institutional investors to monitor their equity trading costs. Its clients include many pension funds and asset managers. The authors do not have permission to redistribute them, even in aggregate form. Requests for this commercial dataset can be addressed directly to the data vendor.  See {\tt www.ancerno.com} for details.

\clearpage

\bibliographystyle{unsrt}
\bibliography{biblio}

\clearpage

\appendix
\section{Statistics of metaorder sample}
\label{app_data}
\begin{figure}[h!]
\centering
\includegraphics[width=0.9\linewidth]{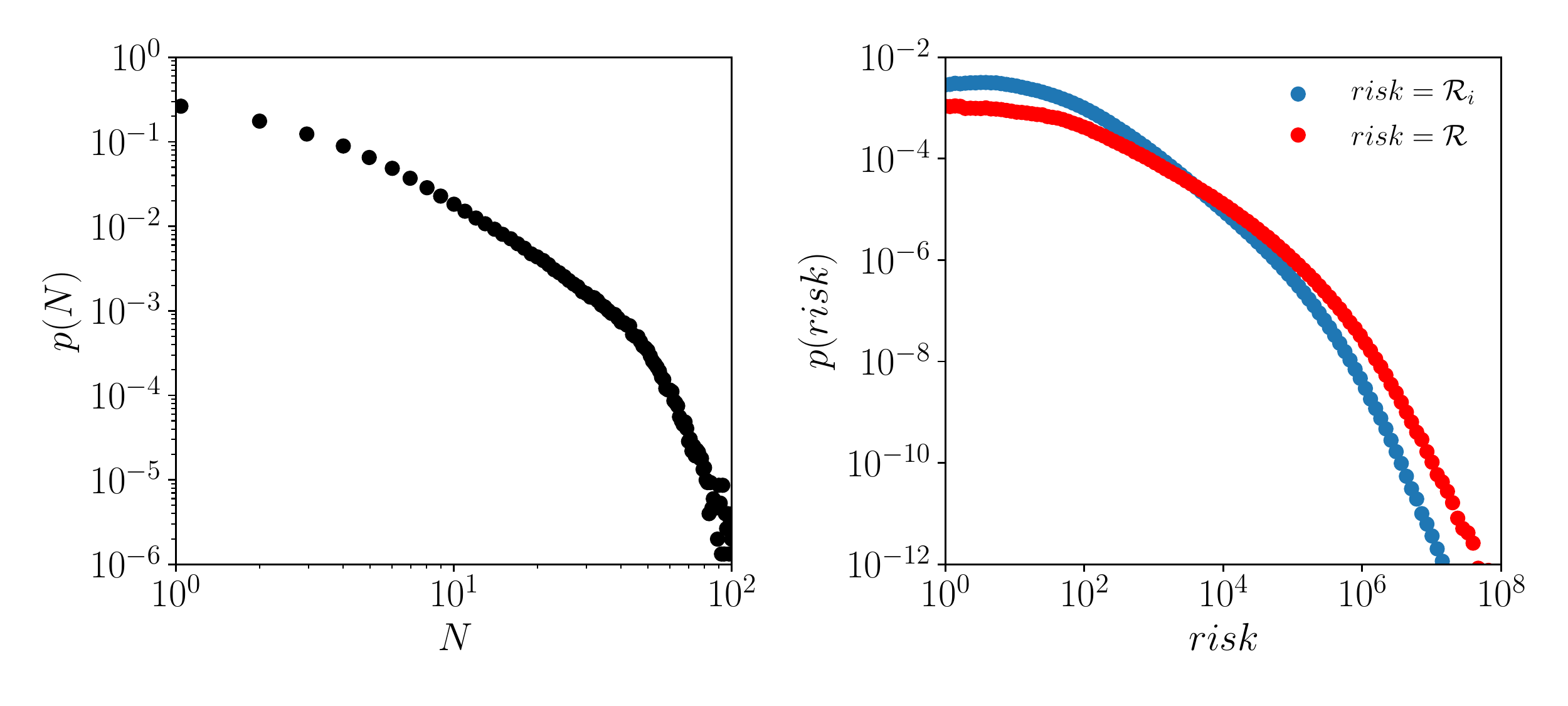}
\caption{(Left panel) Empirical probability distribution of the daily number $N$ of  metaorders per asset: $N$ is broadly distributed over two decades with an average close to 5. (Right panel) Empirical probability distributions of the exchanged risk per metaorder, i.e $\ri_i:=\sigma_{\mathrm{d}}\textrm{v}_i \textrm{p}_i$, and of the total daily risk per day/assets, i.e $\mathcal{R}:=\sum_{i=1}^N \ri_i$.}
\label{fig0}
\end{figure}

Here we  describe  some statistics of the metaorders executed from the main investments funds and brokerage firms  gathered by  ANcerno. The empirical probability distribution of the number of metaorders $N$ per asset, of the risk $\mathcal{R}_i$ exchanged by a metaorder and of the total daily traded risk $\mathcal{R}$ per asset are illustrated in Figure \ref{fig0}. It emerges  that both the number of daily metaorders $N$ and the risk measures typically vary  over several orders of magnitude. In particular, as evident from the left panel in Fig. \ref{fig0},  there is a significant number of metaorders active every day, since in average $\sim$5 metaorders are executed per day for each asset. Furthermore, as shown in the right panel of Fig. \ref{fig0}, both the single bet's risk $\mathcal{R}_i$ and the total daily exchanged risk $\mathcal{R}$ vary  over almost eight decades. Note that these statistical properties are approximately independent from  the time period and from the economical sector of the asset exchanged through metaorders.

\section{The 3/2-law under the microscope}
\label{app_micro}

\begin{figure}[t!]
\centering
\includegraphics[width=0.9\linewidth]{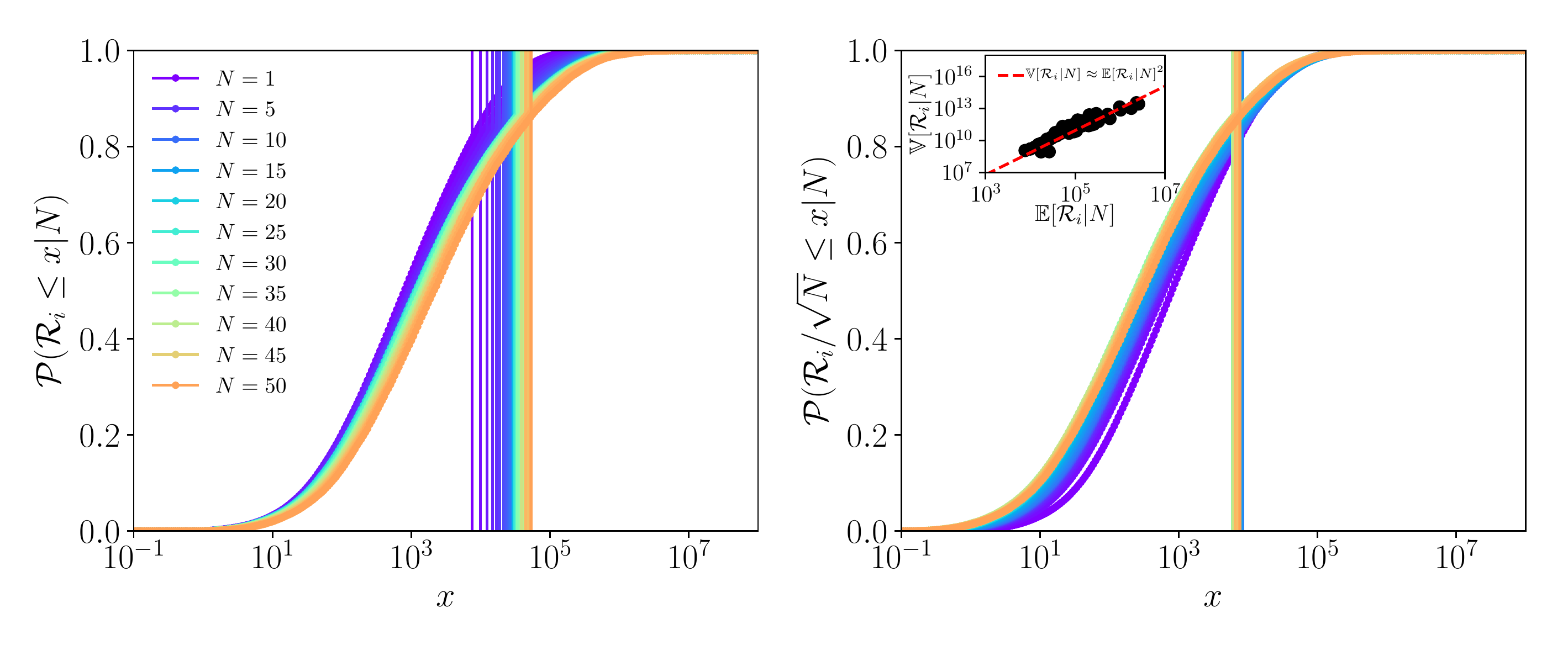}
\caption{Empirical cumulative distribution  of the traded metaorder's risk  $\mathcal{R}_i=\sigma_{\mathrm{d}} \textrm{v}_i\textrm{p}_i$ without (left panel) and with (right panel) rescaling by the square root of the daily number $N$ of  metaorders per asset.  The colored vertical lines represent the location of the average for each  sample conditional on $N$. To note that  also if the empirical distribution is not an invariant function of $N$, we observe that $\avg[\mathcal{R}_i/\sqrt{N}|N]\simeq \textrm{const.}$, as evident from the vertical lines in the right panel,  which is at the origin of the measured 3/2-law.}
\label{figure5}
\end{figure}

\begin{figure}[ht!]
\begin{minipage}{1.0\textwidth}
\centering
\includegraphics[width=0.8\linewidth]{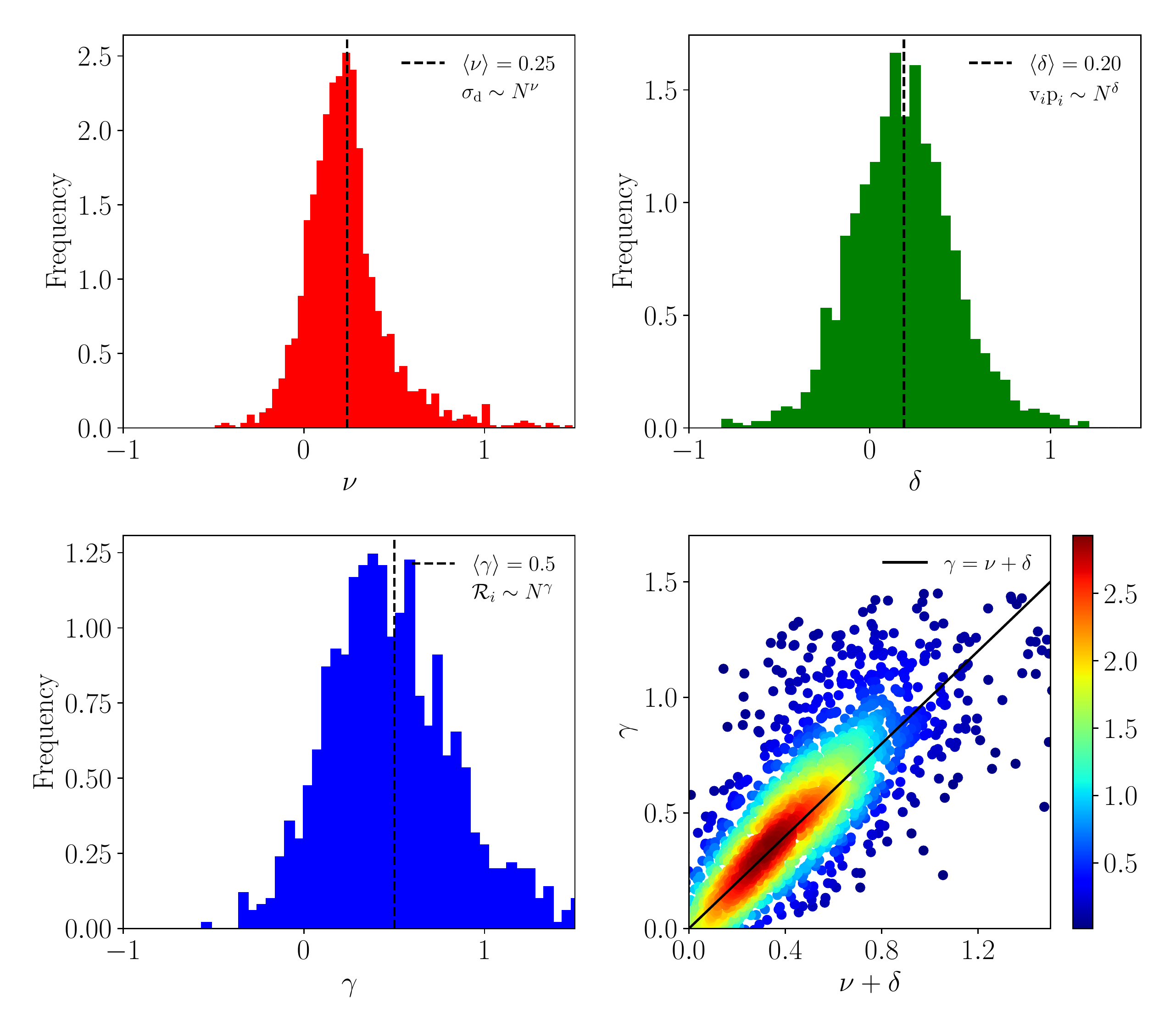}
\end{minipage}
\caption{(Top left panel) Empirical distribution of the scaling exponent $\nu$ computed for each stock regressing $\sigma_{\mathrm{d}} \sim N^{\nu}$: in average $\langle \nu \rangle =0.25$ as shown by the dashed black line. (Top right panel) Empirical distribution of the scaling exponent $\delta$ computed for each stock regressing $\textrm{v}_i \textrm{p}_i \sim N^{\delta}$: in average $\langle \delta \rangle =0.20$ as shown by the dashed black line. (Bottom left panel) Empirical distribution of the scaling exponent $\gamma$ computed for each stock regressing $\ri_i \sim N^{\gamma}$: in average $\langle \gamma \rangle =0.5$ as shown by the dashed black line. (Bottom right panel) Signature scatter plot (coloured by density of data) of the coefficients $\nu+\delta$ and   $\gamma$   respectively estimated conditioning to each stock.}
\label{figureHistCorr}
\end{figure}

One may rightfully wonder whether it is possible to understand the $3/2$-law from the statistical properties of the metaorders.
To this purpose we start by  investigating the individual metaorder's risk $\ri_i$ distribution properties as a function of $N$. We find that
 when rescaling the metaorder's risk $\ri_i$ by the square root of the number $N$ of daily metaorders per asset one obtains a conditional cumulative distribution $\mathcal{P}(\ri_i/\sqrt{N}|N)$ dependent on  $N$ but with a mean $\avg[\mathcal{R}_i/\sqrt{N}|N]$  invariant on $N$ (see Fig.~\ref{figure5}). It emerges then that  the conditional average metaorder risk $\ri_i$ can be predicted from the number $N$ of daily metaorders per asset since $\avg[\mathcal{R}_i|N]$ scales as $N^{\gamma}$ with $\gamma \simeq 0.5$, that is $\avg[\mathcal{R}_i|N]\sim \sqrt{N}$.\footnote{In analogy, the variance $\mathbb{V}[\mathcal{R}_i|N]$ scales linearly with $N$, i.e. $\mathbb{V}[\mathcal{R}_i|N] \approx \avg[\mathcal{R}_i|N]^2 $.}
It immediately follows that combining this empirical result  and  the linearity property of the mean, one recovers the $3/2$-law $\avg[\ri|N]  \sim N^{3/2}$, since: 
\begin{equation}
\avg[\ri|N]=\avg\bigg[ \sum_{i=1}^N \ri_i\Big|N\bigg]=\sum_{i=1}^N\avg[\ri_i|N]=N \hspace{0.09cm}\avg[\ri_i|N]\sim N \sqrt{N}= N^{3/2}.
\label{pippo}
\end{equation}
{To explain the scaling $\avg[\ri_i|N]\sim \sqrt{N}$ through the product $\avg[\sigma_{\mathrm{d}}|N]\times \avg[\textrm{v}_i\textrm{p}_i|N]$ we need to check for the correlation }between the daily volatility $\sigma_{\mathrm{d}}$ and the volume in dollars $\textrm{v}_i\textrm{p}_i$ of a metaorder, which is found to be $\langle \mathbb{C}(\sigma_{\mathrm{d}},\textrm{v}_i \textrm{p}_i)\rangle \approx 3\times10^{-2}$. For each stock we regress $\ri_i \sim N^{\gamma}$, $\sigma_{\mathrm{d}} \sim N^{\nu}$, $\textrm{v}_i\textrm{p}_i \sim N^{\delta}$, {and we obtain} from  the empirical distributions of the exponents in Fig.~\ref{figureHistCorr} that their average values read $\langle \gamma \rangle =0.5$, $\langle \nu \rangle =0.25$ and $\langle \delta \rangle=0.20$, thus $\langle \gamma \rangle \neq  \langle \nu \rangle +\langle \delta \rangle$. 
{However, by looking at the scatter plot of the estimated exponent} $\gamma$ as function of the sum $\nu+\delta$ computed separately for each stock  (see bottom right panel in  Fig.~\ref{figureHistCorr})
one observes a clear linear relation.

A possible and intuitive explanation of the non null measured correlation between $\sigma_{\mathrm{d}}$ and  $\textrm{v}_i\textrm{p}_i$ is that  metaorders add up to volume, generate market impact and thus increase price volatility. In this way trading volume increases due to both an increase in the number of \textit{bets} and in their sizes, and so does volatility from the increased market impact as discussed  for example in  \cite{TVV}.  Note that this reasoning is  valid even if the metaorders  only account for a certain percentage of the total daily market volume $V=\sum_{i=1}^N\textrm{v}_i=\eta V_{\mathrm{d}}$ with  $\eta$ adjusting for the partial view of the ANcerno sample in terms of volume, and for the non-\textit{bet} traded by intermediaries: from our dataset we measure in average $\langle \eta \rangle \approx 5 \times 10^{-2}$.

 \section{Statistics of trading costs}\label{app_costs}
\begin{figure}[ht!]
\begin{minipage}{1.0\textwidth}
\centering
\includegraphics[width=0.85\linewidth]{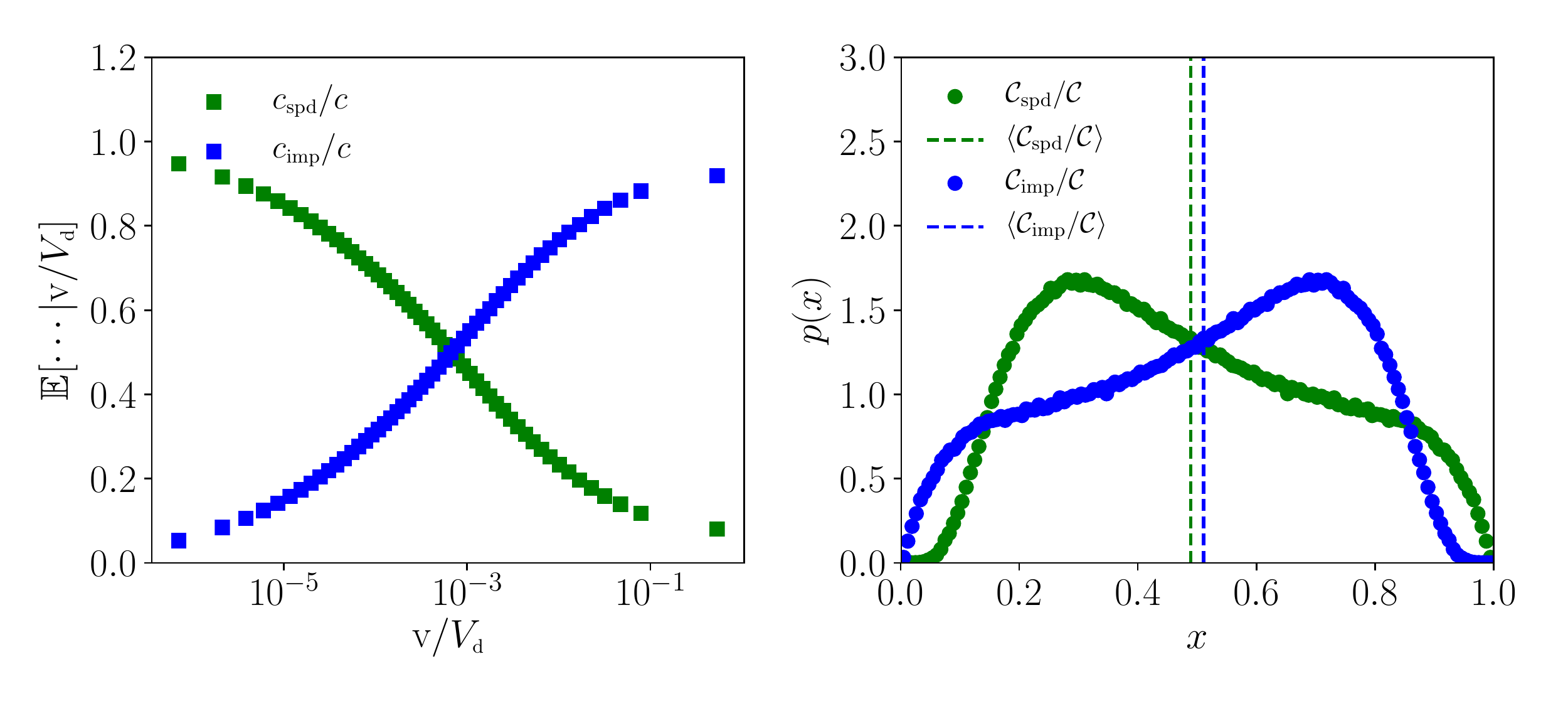}
\end{minipage}
\caption{(Left panel) Averaged spread and market impact  cost ratios given respectively by $c_{\textrm{\mathrm{spd}}}/c$ and $c_{\textrm{imp}}/c$ - with $c_{\textrm{\mathrm{spd}}}=S\times \textrm{v}$ (spread cost), $c_{\textrm{imp}}=\sigma_{\mathrm{d}}\times \textrm{v}\textrm{p}\times \sqrt{\textrm{v}/V_{\mathrm{d}}}$ (market impact cost) and $c=c_{\textrm{\mathrm{spd}}}+c_{\textrm{imp}}$ (total cost per \textit{bet}) -  as function of the metaorder's order  size $\textrm{v}/V_{\mathrm{d}}$:  to note that for a \textit{bet} with small (large) order size the spread (market impact)  cost is dominant. (Right panel) Empirical distributions of the $\mathcal{C}_{\textrm{\mathrm{spd}}}/\mathcal{C}$ and $\mathcal{C}_{\textrm{imp}}/\mathcal{C}$  ratios which give us  an idea of the order of magnitude of the  different  contributions  to the total daily average cost per \textit{bet}  $\mathcal{C}=\mathcal{C}_{\textrm{\mathrm{spd}}}+\mathcal{C}_{\textrm{imp}}$ (computed from Eq. \ref{gencost} fixing $Y_{\textrm{spd}}=3.5$ and $Y_{\textrm{imp}}$=1.5): the dashed vertical lines represent the location of the mean values equal respectively to $\langle \mathcal{C}_{\textrm{\mathrm{spd}}}/\mathcal{C} \rangle=0.49$ and $\langle \mathcal{C}_{\textrm{imp}}/\mathcal{C} \rangle=0.51$. }
\label{figureICItilde}
\end{figure}

 As expected, we find that, for a single \textit{bet} with unsigned volume $\textrm{v}$, the spread cost $c_{\textrm{\mathrm{spd}}}=S\times \textrm{v}$ is dominant for small volumes, while the market impact cost $c_{\textrm{imp}}=\sigma_{\mathrm{d}}\times \textrm{v}\textrm{p}\times\sqrt{\textrm{v}/V_{\mathrm{d}}}$  takes over for large volumes (see left panel of Fig.~\ref{figureICItilde}).  Furthermore, as shown in the right panel of Fig.~\ref{figureICItilde},  the average daily market impact cost $\mathcal{C}_{\textrm{imp}}$  accounts on average for $\approx 1/2$ of the total daily trading average cost  $\mathcal{C}=\mathcal{C}_{\textrm{\mathrm{spd}}}+\mathcal{C}_{\textrm{imp}}$, computed using $Y_0=3.5$ and $Y=1.5$ in Eq.~\eqref{gencost}.

\end{document}